\documentclass{pas}

\usepackage{multirow}
\usepackage{comment}
\usepackage{mathtools}
\usepackage{hyperref}
\hypersetup{
    colorlinks=true,
    linkcolor=blue,
    citecolor=blue,
    urlcolor=green,
    }
\usepackage{graphicx}
\usepackage{txfonts}
\usepackage{ulem}
\usepackage{threeparttable}
\usepackage{mathrsfs}
\usepackage{amsmath}
\usepackage{placeins}
\usepackage{afterpage}
\usepackage{caption}
\usepackage{rotating}
\begin{document}

\lefttitle{Publications of the Astronomical Society of Australia}
\righttitle{L. Bruno et al.}

\jnlPage{1}{4}
\jnlDoiYr{2021}
\doival{10.1017/pasa.xxxx.xx}

\articletitt{Research Paper}

\title{The Northern Cross Fast Radio Burst project: VI. The \textit{INCART} public database}

\author{\sn{L.} \gn{Bruno}$^{1}$, 
\sn{G.} \gn{Bernardi}$^{1,2,3}$, 
\sn{M.} \gn{Pilia}$^{4}$, 
\sn{D.} \gn{Pelliciari}$^{1}$, 
\sn{A.} \gn{Geminardi}$^{5,6,4}$, 
\sn{F.} \gn{Fiori}$^{1}$,
\sn{V.} \gn{Galluzzi}$^{1,7}$, 
\sn{G.} \gn{Naldi}$^{1}$, 
\sn{M.} \gn{Trudu}$^{4}$,
\sn{A.} \gn{Zanichelli}$^{1}$} 

\affil{$^1$Istituto Nazionale di Astrofisica (INAF) - Istituto di Radioastronomia (IRA), via Gobetti 101, 40129 Bologna, Italy. $^2$ South African Radio Astronomy Observatory, Black River Park, 2 Fir Street, Observatory, Cape Town, 7925, South Africa. $^3$ Department of Physics and Electronics, Rhodes University, PO Box 94, Makhanda, 6140, South Africa. $^4$ Istituto Nazionale di Astrofisica (INAF) - Osservatorio Astronomico di Cagliari (OAC), via della Scienza 5, I-09047, Selargius (CA), Italy. $^5$ Scuola Universitaria Superiore IUSS Pavia, Palazzo del Broletto, piazza della Vittoria 15, I-27100 Pavia, Italy. $^6$ Dipartimento di Fisica, Università di Trento, via Sommarive 14, I-38123 Povo (TN), Italy. $^7$ Istituto Nazionale di Astrofisica (INAF) - Italian Centre for Astronomical Archives (IA2), via G.B. Tiepolo, 11, I-34143 Trieste, Italy. }

\corresp{L. Bruno, Email: l.bruno@ira.inaf.it}

\citeauth{L. Bruno, G. Bernardi, M. Pilia, D. Pelliciari, A. Geminardi, F. Fiori, V. Galluzzi, G. Naldi, M. Trudu, A. Zanichelli (2026), The Northern Cross Fast Radio Burst project: VI. The INCART
public database. {\it Publications of the Astronomical Society of Australia} {\bf 00}, 1--12. https://doi.org/10.1017/pasa.xxxx.xx}

\history{(Received xx xx xxxx; revised xx xx xxxx; accepted xx xx xxxx)}

\begin{abstract}
Fast radio bursts (FRBs) are bright (Jansky-level) and short-duration ($\sim 1$  ms) flashes of extragalactic origin. 
Observations of single events have now been complemented by large-area surveys, delivering FRB catalogues and enabling the first population studies. The Northern Cross (NC) radio interferometer 
is one of the instruments performing observations of FRBs. In this work, we present the Italian Northern Cross Atlas of Radio Transients ({\tt INCART}), a public platform for the distribution of data products from the NC. {\tt INCART} makes available to the community the FRBs observed by the NC through manageable frequency-time series datasets and catalogues with best-fit physical parameters. The design of {\tt INCART} guarantees the possibility of scientific re-analysis of the FRB properties, in view also of future releases of the processing pipeline. Furthermore, {\tt INCART} focuses on long-term storage optimisation, which is a key aspect of state-of-the-art instrumentation. Public access to the FRB data from the NC maximises the legacy value of the collection, facilitates the synergy with other publicly-available catalogues, and fosters research group collaborations.

\end{abstract}


\maketitle

\section{Introduction}

Fast radio bursts (FRBs) are bright (Jansky-level) flashes of short duration ($\sim 1$  ms). They were first discovered in 2007 \citep{lorimer07} and nowadays thousands of FRB events have been reported \citep{CHIME26CAT2}. As a consequence of the propagation of the signal through an ionised medium, radio waves undergo dispersion, causing a delay of the arrival time at lower frequencies. This delay depends on the electron density along the path between the source and the observer, and is parametrised by the dispersion measure ${\rm DM}\propto \int n_{\rm e}dl$. FRBs exhibit high DM values, which are incompatible with the dispersion caused solely by the Milky Way medium, and localisation studies have indeed confirmed that their origin is extragalactic (e.g. \citealt{Tendulkar17,Bannister19}). However, the progenitor and emission mechanisms producing FRBs are still unclear (e.g. \citealt{Cordes&Chatterjee19,Petroff19,Bailes22,Zhang23FRB}, for reviews).  It is well established that a fraction of FRBs show repeating burst activity \citep[e.g.][]{Spitler16,CHIME/FRBCollaboration19,CHIME26CAT2}, but it is unclear whether all FRBs do repeat and if repeating and non-repeating FRBs arise from the same processes \citep[e.g.][]{Chime/FrbCollaboration23,Kirsten24}. Owing to their extragalactic origin and diverse intervening environments they propagate through, FRBs are useful cosmological probes \citep[e.g.][]{Macquart20,Bhandari21,Wu22,James22,Zhang23a,Zhao23}. Therefore, delivering catalogues of FRBs is fundamental to investigate their origin, plan follow-up observing campaigns, and constrain cosmological parameters.

The Northern Cross (NC; \citealt{Bianchi23}) is a T-shaped transit interferometer consisting of two perpendicular arms oriented to N-S and E-W, built in the 1960s in Medicina ($\sim 30$ km to Bologna, Italy). In the past, it was used to scan the sky at 408 MHz and delivered catalogues of extragalactic radio sources \citep[e.g.][]{Colla70,Fanti74,Pedani&Grueff99}. In the recent years, the NC has undergone a major upgrade that enabled the detection of FRBs (see \citealt{locatelli20} and references therein). With this upgrade, the NC is becoming a leading facility in Europe for the observation of FRBs. Since the first FRB detection with the NC in 2021 \citep{trudu22}, a total of 31 events from 6 different FRBs have been detected as of the end of 2025 (\citealt{trudu22,pelliciari23,pelliciari24}; Geminardi et al. in prep.), demonstrating the capability of the instrument. With the completion of the NC refurbishment, the {\it Next Generation Croce del Nord} (NG-Croce) will have improved capabilities, with higher sensitivity, possibility of performing multi-beam observations, and a transient buffer for real-time detection \citep{Naldi25}.

The synergy between the NC and other national and international FRB detectors is becoming relevant for follow-up monitoring, localisation, spectral studies, and statistical analyses (e.g. \citealt{Pilia20,Casentini25,Shah25}). In this paper, we present the Italian Northern Cross Atlas for Radio Transients ({\tt INCART}\footnote{\url{https://ngc-frb-incart.ira.inaf.it/}}), a public database that provides the scientific community with access to the Northern Cross -- fast radio burst (NC--FRB) data products.

The paper is organised as follows. In Section \ref{sect: Data processing}, we summarise typical observation setup, data processing, and analysis for NC--FRB data. In Section \ref{sect: The INCART system}, we present {\tt INCART} and describe the procedures leading to the archived data products. In Section \ref{sect: Conclusions}, we discuss the key issue of large data volumes, and how this is handled by the NC and other FRB detectors. In Section \ref{sect: Conclusions2} we summarise our work and the prospects for the NG-Croce.

\section{Observations and data processing}
\label{sect: Data processing}

\begin{table}
\centering
	\caption[]{Properties of current NC filterbank-format data.}
	\label{table: NC par}   
	\begin{tabular}{cccc}
	\hline
	\noalign{\smallskip}
Parameter & Value & Units & Description \\ 
\hline
$\nu_{\rm min}$  & 401.039 & MHz & Last frequency channel \\
$\nu_{\rm max}$  & 415.854 & MHz & First frequency channel\\
$\nu_{\rm c}$  & 408.447 & MHz & Central frequency channel\\
$\Delta\nu_{\rm chan}$ & 14.468 & kHz & Channel width \\
$N_{\rm chan}$ & 1024 & & Number of channels \\ 
BW & 14.8 & MHz & Total bandwidth \\
$N_{\rm pol}$ & 1 & & Number of polarisations \\ 
$\tau_{\rm samp}$ & 138.24 & $\mu {\rm s}$ &  Temporal resolution \\ 
\noalign{\smallskip}
	\hline
	\end{tabular}  
\end{table}

The NC operates in P-band, with a central frequency of 408 MHz and an effective bandwidth of $\sim 15$ MHz. In the current configuration, only the N-S arm is working, whereas the E-W arm is under refurbishment. The N-S arm includes 64 reflective parabolic cylinders\footnote{Observations carried out before 21-Mar.-2021 employed only 6 out of 64 total cylinders. Afterwards, 8 cylinders were used during 2021-2022. Starting from 2023, 16 cylinders have been used, while the remaining cylinders are currently under refurbishment.}, each one focusing the incoming radiation towards groups of dipoles along the focal line. As detailed in \cite{locatelli20}, the acquired signal is digitised and processed. The processing chain includes correction for cable mismatch, data channelisation, correction for delays, and combination of the response of the dipoles into a single beam. The resulting beam-formed data are converted into total intensity data stored in a custom time-frequency format, which have a typical size of $\sim 250$ GB as obtained from an observing run of 1 hour with the present NC working configuration. Finally, these data are equalised and rescaled  using {\tt digifil} from the {\tt DSPSR} library \citep{VanStraten&Bailes11}, converting them into a standard {\tt SigProc} \citep{Lorimer11} filterbank format suitable for transient search analysis. Currently, the NC--FRB filterbank-format data are saved with 1024 frequency channels of width 14.468 kHz each and a temporal resolution of $138.24 \; \mu{\rm s}$ (see Table \ref{table: NC par}). With this procedure, the data size is reduced to $\sim 50$ GB.

The antenna gain calibration is performed by deriving corrections from the observation of a bright calibrator at transit, with Cassiopeia A (Cas A) routinely used for the NC. Although the Radio Frequency Interference (RFI) level is typically low, various tools can be used to excise the bulk of RFI, such as {\tt RFIFIND} \citep{Ransom02} and {\tt IQRM} \citep{Morello22}. De-dispersion trials are performed over wide DM ranges to search for possible transients through {\tt HEIMDALL} \citep{Barsdell12}. This provides estimates of the arrival time, duration (as boxcar width), signal-to-noise ratio (SNR), and DM of the candidates. The candidate events are validated by {\tt FETCH} \citep{Agarwal20}, which employs convolutional neural networks trained to distinguish spurious from genuine signals.

\section{The \textit{INCART} database}
\label{sect: The INCART system}

\begin{figure*}
        \centering

\includegraphics[width=0.8\textwidth]{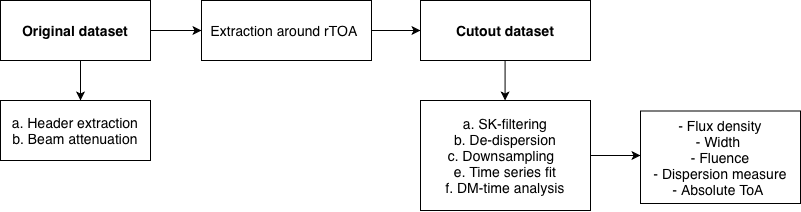}

        \caption{Flowchart of data analysis steps performed by the NC {\tt archivist} pipeline (Section \ref{sect: Re-analysis steps}).  }
        \label{fig: steps}
\end{figure*}

\begin{figure}
        \centering
\includegraphics[width=0.4\textwidth]{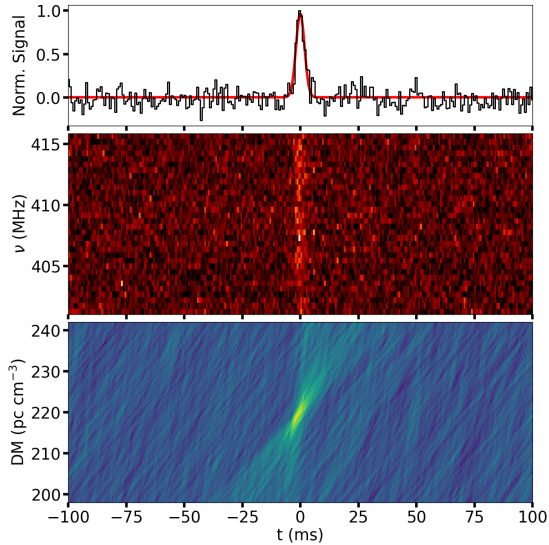}

        \caption{Example of diagnostic plots for FRB20220912A (observation of 26-Aug.-2023, see Table \ref{table: frb ex}). {\it Top}: De-dispersed downsampled time series normalised to the burst peak. The red line shows the Gaussian fit to the burst profile. {\it Centre}: Dynamic spectrum. For clearer inspection, a  downsampling of 32 frequency channels is applied. {\it Bottom}: Butterfly diagram showing the distribution of the time series around different DM values. }
        \label{fig: frb ex}
\end{figure}

\begin{table}
 \fontsize{7}{7}\selectfont
	\caption[]{Example of main entries in the logfile for FRB20220912A (observation of 26-Aug.-2023, see Figure \ref{fig: frb ex}).}
	\label{table: frb ex}   
	\begin{tabular}{cc}
	\hline
	\noalign{\smallskip}
Source Name & FRB20220912A \\
Source RA (J2000) & 23:09:04.9 \\
Source DEC (J2000) & +48:42:25.4 \\
Time stamp of first sample (MJD) & 60182.077245883971 \\
Observation length (minutes) & 34.1 \\
Original relative TOA (s) & 786.5767 \\
Time stamp of first sample cutout (MJD) & 60182.086338206886  \\
Beam correction factor & 1.036 \\
Cutout relative TOA (s) & 1.0 \\
SK masking fraction & 0.00098 \\
Downsampled bins & 256 \\
Temporal window (ms) & 200 \\
Downsampled time ($\mu$s) & 781.25 \\
Norm. Gauss Amp. & 0.98 \\
Err. Norm. Gauss Amp. & 0.06 \\
Fit Width (ms) & 4.3 \\
Err. Fit Width (ms) & 0.3 \\
Fit gauss center (ms wrt TOA) & -0.035 \\
SNR peak & 10 \\
SNR integrated & 25 \\
Gauss fit Chi2red & 1.04 \\
Flux density peak (Jy) & 12.8 \\
Err. Flux density peak (Jy) & 0.8 \\
Rectangular Fluence (Jy ms) & 54 \\
Err. Rectangular Fluence (Jy ms) & 5 \\
Gaussian Fluence (Jy ms) & 55 \\
Err. Gaussian Fluence (Jy ms) & 5 \\
Fit DM (pc cm-3) & 220 \\
Err. Fit DM (pc cm-3) & 2 \\
Topocentric TOA at chan1 (UTC) & 2023-08-26 02:04:20.621 \\
Topocentric TOA at infinite frequency (UTC) & 2023-08-26 02:04:15.343 \\
Topocentric TOA at infinite frequency (MJD) & 60182.086288693856 \\
\noalign{\smallskip}
	\hline
	\end{tabular}  
\end{table}

We aim to provide the community with access to the NC--FRB data. Raw voltages and beam-formed data are necessary for independent reprocessing, especially to recover potential bursts that may have been undetected due to limitations of the current pipelines. In practice, in the perspective of data storage and in view of the NG-Croce upgrade, preserving and distributing these data for both FRB detections and non-detections is unsustainable in the long term. On the other hand, a catalogue reporting physical quantities derived from the burst analysis is essential, but alone it does not enable independent verification of these values. 

The {\tt INCART} database serves both as a catalogue of FRB properties and as a repository of short time chunks of fully processed filterbank datasets containing the FRB event. This design minimises storage requirements, ensures long-term scalability, provides science-ready data products, and offers the possibility for users to analyse the FRB properties with independent methods. {\tt INCART} is tailored for the NC--FRB project and provides a basic querying system, with search options by FRB name and observation date. In contrast, the \textit{Istituto Nazionale di Astrofisica} (INAF) Radio Data Archive\footnote{\url{http://radioarchive.inaf.it/}} is the national infrastructure that archives and distributes data from all Italian radio facilities, developed prior to and independently from {\tt INCART} for general purposes. For consistency between the two services, the same cutout filterbank datasets available in {\tt INCART}, accompanied by metadata for instrumental, observational, and provenance information, will also be accessible in the Radio Archive, which offers a more advanced and flexible querying system.

The products retrievable via {\tt INCART} are the result of the NC {\tt archivist} pipeline, which was developed for the specific purpose of preparing sets of data curated with homogeneous methods. Therefore, {\tt archivist} was also run on previously published NC--FRB data \citep{trudu22,pelliciari23,pelliciari24} for consistency with the whole collection. In Sections \ref{sect: Re-analysis steps}, \ref{sect: Equazioni} we describe the implemented procedure.

\subsection{FRB analysis steps}
\label{sect: Re-analysis steps}

The {\tt archivist} v. 1.0 pipeline requires the processed filterbank dataset (Section \ref{sect: Data processing}) containing the confirmed FRB event, which we will refer to as the ‘original’ dataset. The fundamental input parameters, which can be obtained for example from {\tt HEIMDALL}, are i) the time of arrival relative to the observation start (rToA), which is referred to the maximum frequency channel ($\nu_{\rm max}$, Table \ref{table: NC par}), ii) width ($W$), which is the FWHM of the burst profile, and iii) DM. The input values of $W$ and DM need to be representative priors, but they are then refined by the analysis. In the following, we describe the steps displayed in the flowchart of Figure \ref{fig: steps}, which lead to the final data products accessible in {\tt INCART}. The detailed calculations performed by {\tt archivist} are reported in Section \ref{sect: Equazioni}.

\begin{enumerate}
    \item Metadata from the header of the original dataset are saved in a logfile. The logfile will be progressively updated with additional metadata that are not included in the header and results of the FRB analysis.  
    \item The sensitivity of transit telescopes  towards a specific direction varies in time during an observing run. The attenuation of the signal needs to be corrected by a factor $f_{\rm beam}({\rm rToA})$, which is computed from the total observing time and the NC beam parameters (see e.g. \citealt{Geminardi25} for details).
    \item Based on the input $W$ and DM values, the original dataset is cut out in time around rTOA. This ‘cutout' dataset contains a few seconds of observation centred on the FRB event, typically reducing the data size from $\sim 50 $ GB to $\sim 30$ MB, and enabling sustainable preservation and easy distribution. Subsequent steps will be performed on the cutout dataset. 
    \item Possible RFI is identified using a spectral kurtosis filter \citep{Vrabie03,Nita&Hellbourg19}, and individual time–frequency samples exceeding a $5\sigma_{\rm SK}$ threshold are masked. Given that the analysis is performed on a short cutout centred on the burst, this approach is adequate and does not require flagging of entire frequency channels or time intervals. The typical fraction of masked samples is $\xi_{\rm SK}\lesssim 1\%$. 
    \item Data are de-dispersed around the input DM. A proper downsampling is then applied taking into account the input burst width and cutout length, preserving a sufficient number of temporal bins to fit the FRB profile and estimate the noise level. Specifically, downsampling is performed with the {\tt local\_resize\_mean} function within the {\tt scikit-image} python package \citep{vanderWalt14}, yielding to a number of temporal bins scaling as a power of 2 within a narrow (a few hundreds of ms) time window. Such procedure optimises the SNR, while typically decreasing the temporal resolution from $\sim 0.1$ ms to $\sim 1$ ms. 
    \item All frequency channels are averaged to obtain the de-dispersed time series, which is then normalised by the peak value for clearer inspection. The time series is fitted with a Gaussian function, providing best-fit values of its centre, amplitude ($A$), and standard deviation ($\sigma$). The obtained $A$ and $\sigma$ are used to compute the SNR, which provides flux density and fluence measurements after conversion from instrumental to physical units. 
    \item The representative DM used in previous steps is the result of FRB searching algorithms (Section \ref{sect: Data processing}) and is thus adequate for the burst characterisation. For refinement and uncertainty estimation, values within the range ${\rm DM} \pm 0.1\times {\rm DM}$ are tested. This range is sampled with 256 steps (providing sufficient DM resolution, while keeping the computational cost manageable), and the data are de-dispersed at each DM. We consider the DM that maximises the SNR, DM$_{\rm max}$, as the best estimate.   
    \item Using rToA, the absolute topocentric ToA (referred to $\nu_{\rm max}$) is computed. This is scaled to $\nu \rightarrow \infty$ by correcting for the dispersive delay assuming ${\rm DM= DM}_{\rm max}$.    
    \end{enumerate}

The products stored in {\tt INCART} are the cutout filterbank dataset as obtained from step 3, the logfile, and sets of plots showing the normalised de-dispersed time series, dynamic spectrum, and time-DM (‘butterfly') diagram. Examples of the logfile and output plots for the repeating source FRB20220912A (observation date: 26-Aug.-2023) are shown in Table \ref{table: frb ex} and Figure \ref{fig: frb ex}.

\subsection{Procedures and equations}
\label{sect: Equazioni}

\begin{figure}
        \centering
\includegraphics[width=0.4\textwidth]{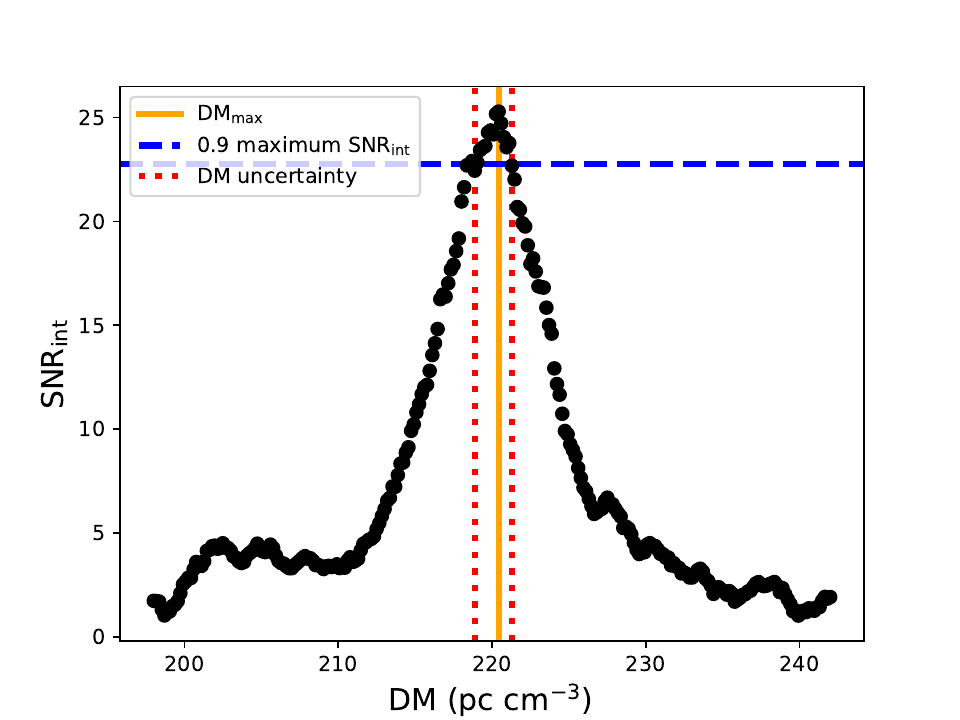}
        \caption{Example of ${\rm SNR_{int}}$-DM distribution for FRB20220912A (Figure \ref{fig: frb ex}, Table \ref{table: frb ex}). The value of DM$_{\rm max}$ (orange solid line) is the DM maximising ${\rm SNR_{int}}$ (step 7 in Section \ref{sect: Re-analysis steps}). The corresponding uncertainty range (red dotted lines) consists of the DM values providing a decrease of the maximum ${\rm SNR_{int}}$ by the $10\%$ (blue dashed line). }
        \label{fig: dm-snr}
\end{figure}

In this Section we report the detailed procedures performed by the NC {\tt archivist} v. 1.0 pipeline following the steps described in Section \ref{sect: Re-analysis steps}. We assume that the FRB time series can be described by a Gaussian profile in the form:
\begin{equation}
    G(t)=Ae^{-\frac{1}{2}\left( \frac{t-{\rm rToA}}{\sigma } \right)^2} \; \; \; ,
    \label{eq: gauss profile}
\end{equation}
where $A$ is the amplitude (in units of ${\rm s}^{-1}$), $\sigma=W/\sqrt{8\ln{2}}$ is the standard deviation (with $W={\rm FWHM}$, in units of ${\rm s}$), and rToA is the time of arrival relative to the observation starting time. Integrating $G(t)$ over time in ${\rm rToA}\pm N\sigma$, the resulting area is:
\begin{equation}
    \mathcal{A}_{\rm Gauss}= \int^{{\rm rToA}+ N\sigma}_{{\rm rToA}- N\sigma} G(t){\rm d}t=A\sqrt{2\pi}\sigma f(N) \; \; \; ,
    \label{eq: gauss area}
\end{equation}
where $f(N)= 1$ for $N\rightarrow \infty$. In our analysis, the time series is normalised to its peak value, implying that the fitted amplitude is expected to be $A\sim 1 \; {\rm s}^{-1}$ for FRBs with a single burst. The goodness of the fit is evaluated through the reduced $\chi^2$. We considered integration limits with $N=2$, thus setting $f(2)\sim 0.95$ in Equation \ref{eq: gauss area} (see below the motivation for this choice). 

The fitted parameters are used to compute the FRB flux density and fluence. The flux density of the peak is obtained as
\begin{equation}
    S_{\rm p} = {\rm SNR}_{\rm p} \times \sigma_{\rm therm}  \; \; \; ,
    \label{eq: flux peak}
\end{equation}
where ${\rm SNR}_{\rm p}=A/\sigma_{\rm std}$ is the ratio of the amplitude to the standard deviation of the noise computed in off-burst bins (that is, beyond the considered integration limits), and $\sigma_{\rm therm}$ is the thermal noise (in units of Jy). The thermal noise is computed from the radiometer equation \citep{Lorimer&Kramer12} as:
\begin{equation}
    \sigma_{\rm therm} =  \frac{{\rm SEFD}}{A_{\rm rec} \sqrt{N_{\rm pol}\times {\rm BW} \times (1-\xi) \times t_{\rm samp}}} \; \; \; ,
    \label{eq: therm noise}
\end{equation}
where ${\rm SEFD}$ is the system equivalent flux density, $A_{\rm rec}=4\times {\rm N}_{\rm cyl}$ is the number of receivers based on the number of working cylinders, $N_{\rm pol}$ is the number of recorded polarisations, ${\rm BW}$ is the bandwidth, $\xi$ is the fraction of masked time-frequency samples (we assumed $\xi=\xi_{\rm SK}$, based on step 4 in Section \ref{sect: Re-analysis steps}), and $t_{\rm samp}$ is the considered sampling time. For the NC, ${\rm SEFD}=8400\pm 420$ Jy \citep{trudu22}, $N_{\rm pol}=1$, and ${\rm BW}=14.8$ MHz (Table \ref{table: NC par}). 

For a comparison with previous works on NC--FRBs \citep{trudu22,pelliciari24,Geminardi25}, we derived the fluence via two different methods. In the first method, the ‘rectangular' fluence is computed as the area of a rectangle having sides  equal to $S_{\rm p}$ and $W$, being:
\begin{equation}
    F_{\rm rect} = S_{\rm p} \times W \; \; \; .
    \label{eq: rect fluence}
\end{equation}
In the second method, we took into account the whole duration of the burst, which provides a more effective SNR estimate, rather than using only the peak. Such SNR is given by:
\begin{equation}
    {\rm SNR}_{\rm int} = \frac{\mathcal{A}_{\rm Gauss}}{\sigma_{\rm std}\sqrt{N_{\rm bin}}t_{\rm samp}}   \; \; \; ,
    \label{eq: integrated SNR}
\end{equation}
where $N_{\rm bin}=W/t_{\rm samp}$ is the number of temporal bins covering the burst FWHM, thus allowing a proper comparison between the FRB and noise level. The corresponding thermal noise, $\sigma_{\rm therm, int}$, is obtained as in Equation \ref{eq: therm noise} by substituting $t_{\rm samp}$ with $W$ to take into account the burst duration. The ‘Gaussian' fluence is thus computed as:
\begin{equation}
    F_{\rm Gauss} = {\rm SNR}_{\rm int} \times \sigma_{\rm therm, int} \times W \; \; \; .
    \label{eq: gauss fluence}
\end{equation}

As a rough estimate, the accuracy in the calculation of the fluence via the two methods described above is provided by the ratio of the rectangular ($\mathcal{A}_{\rm rect}=A\times W$) to Gaussian areas:
\begin{equation}
    \frac{\mathcal{A}_{\rm rect}}{\mathcal{A}_{\rm Gauss}}=\sqrt{\frac{4\ln2}{\pi}} \frac{1}{f(N)} \sim \frac{0.94}{f(N)}\; \; \; .
    \label{eq: area ratio}
\end{equation}
Under the assumption that the profile in Equation \ref{eq: gauss profile} is a good representation of the data, Equation \ref{eq: area ratio} shows that for $N= 2$ the two methods are approximately equivalent. 

The logfile reports the peak flux density and fluences calculated with the rectangular and Gaussian methods as described above. Within {\tt INCART}, these values are multiplied by $f_{\rm beam}({\rm rToA})$ to correct for the beam attenuation (step 2 in Section \ref{sect: Re-analysis steps}). The errors on $S_{\rm peak}$ and $W$ are obtained from the fitting uncertainty on $A$ and $\sigma$, respectively. Following the standard propagation error formula, we computed errors on the fluences.

The DM-time analysis producing the butterfly diagram (step 7 in Section \ref{sect: Re-analysis steps}) shows the signal intensity as a function of time and trial DM values, obtained by de-dispersing the data over a range of DMs. This yields a DM value, ${\rm DM}_{\rm max}$, which maximizes ${\rm SNR_{int}}$. To estimate the uncertainty on ${\rm DM}_{\rm max}$, we considered the distribution of ${\rm SNR_{int}}$, and derived the range of DM where the maximum ${\rm SNR_{int}}$ value decreases by $10\%$\footnote{This threshold is empirical and was found to provide a robust estimate of the DM uncertainty, capturing the width of the peak in the SNR curve and avoiding sensitivity to small-scale fluctuations.} (see Figure \ref{fig: dm-snr}). If such DM range is not symmetric with respect to ${\rm DM}_{\rm max}$, the error is obtained as the largest value between the lower and upper limit. This procedure yields relative DM uncertainties of the order of $\sim 0.1-1\%$.

The obtained ${\rm DM}_{\rm max}$ is used to derive the absolute arrival time of the burst in a topocentric reference system at infinite frequency. In modified Julian day (MJD) units, the absolute ToA at the highest frequency channel ($\nu_{\rm max}$ in Table \ref{table: NC par}) is computed as
\begin{equation}
{\rm ToA}_{\rm \nu_{\rm max}} = \frac{t_{0}}{{\rm MJD}} +  \frac{\rm rToA}{86400} \; \; \; .
    \label{eq: toa416}
\end{equation}
The ToA is finally corrected by the dispersive delay and referred to $\nu \rightarrow \infty$ as
\begin{equation}
{\rm ToA}_{\rm \infty} = \frac{{\rm ToA}_{\nu_{\rm max}}}{{\rm MJD}} - \frac{k}{86400} \times \frac{\rm DM_{max}}{\nu_{\rm max}^2} \; \; \; ,
    \label{eq: toainfinity}
\end{equation}
where $k=4148.808 \; {\rm s\; MHz^2 \; cm^3\; pc^{-1}\;   }$ is the dispersion constant relating the delay with DM and $\nu$.

\subsection{Caveats and improvements}
\label{sect: Caveats}

\begin{figure}
        \centering
\includegraphics[width=0.4\textwidth]{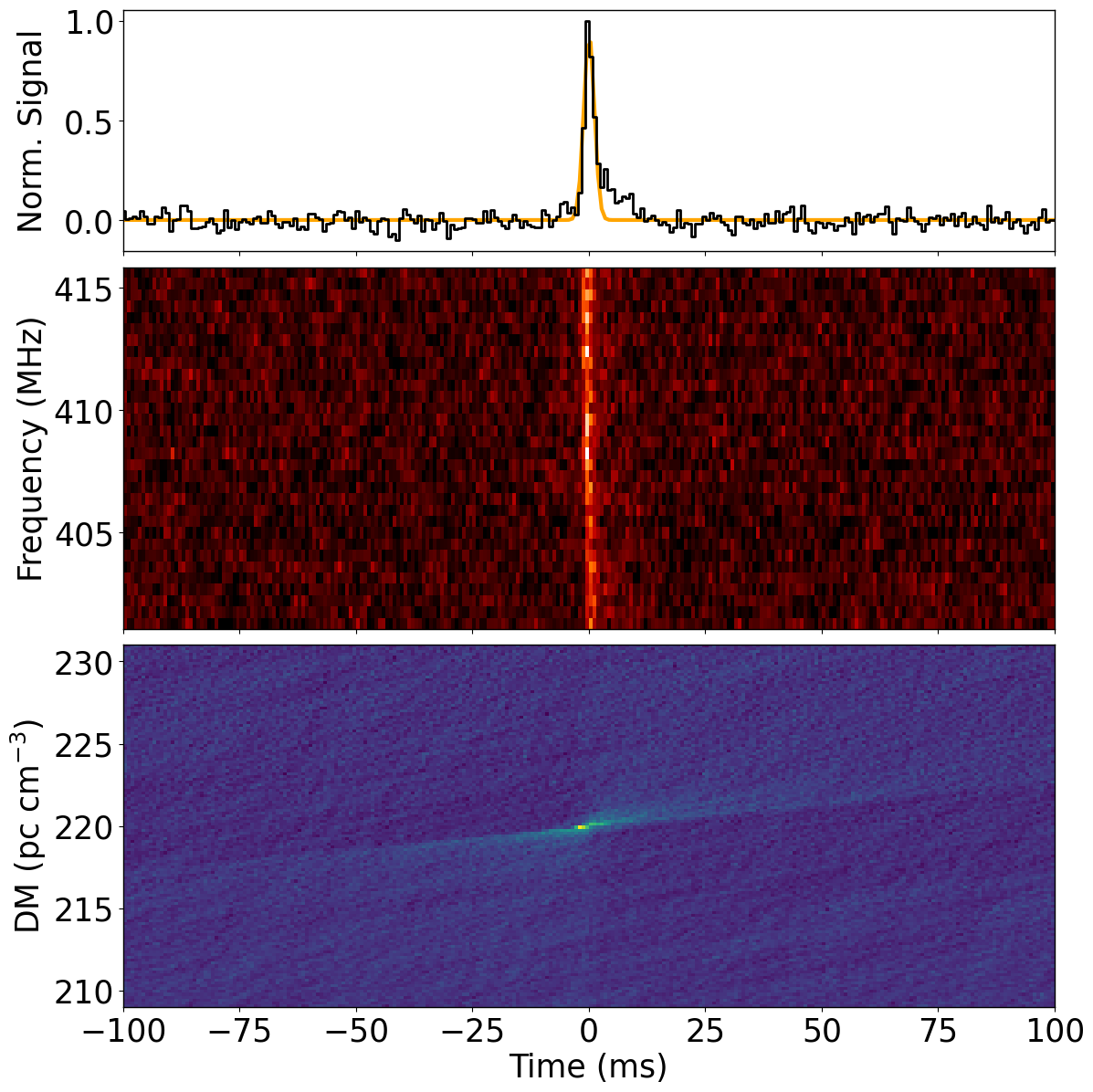}

        \caption{Diagnostic plots for FRB20220912A (observation of 22-Aug.-2023). The FRB profile shows a decaying tail that is not adequately reproduced by the considered Gaussian profile.  }
        \label{fig: frb ex3}
\end{figure}

\begin{figure*}
        \centering
\includegraphics[width=0.4\textwidth]{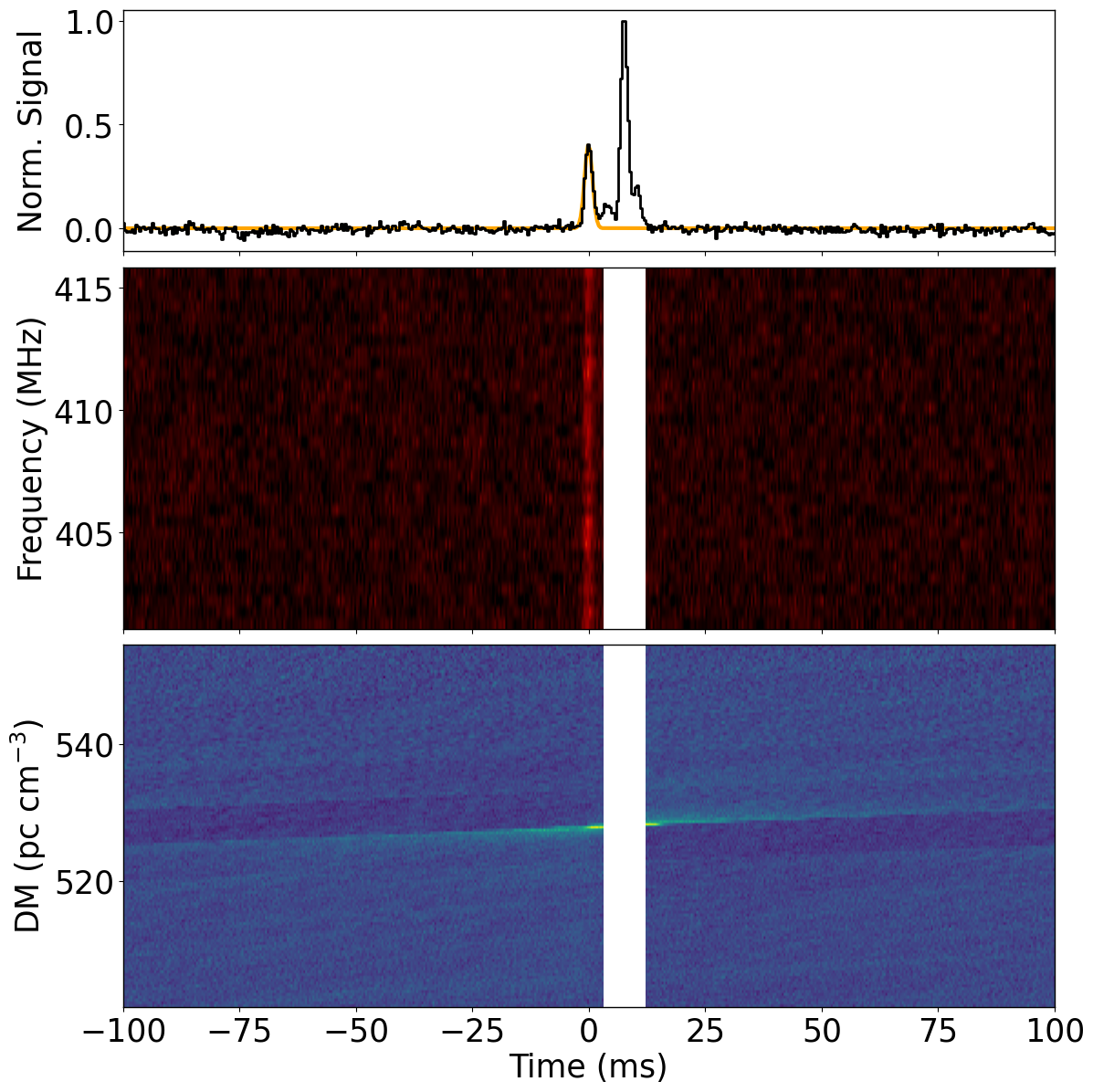}
\includegraphics[width=0.4\textwidth]{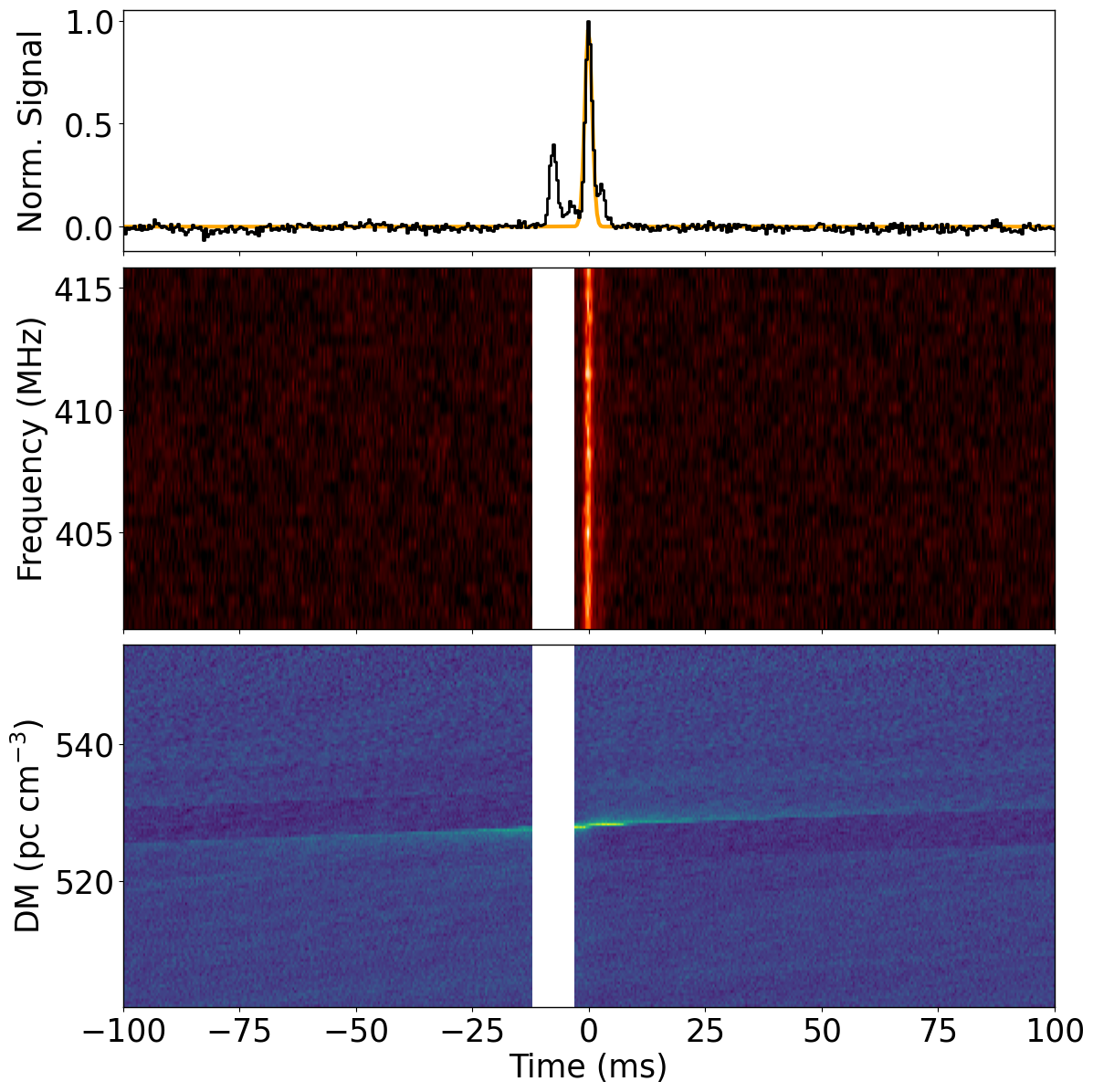}

        \caption{Diagnostic plots for FRB20240114A (observation of 17-Mar.-2024) showing sub-structure in the time series. The two main bursts were fitted separately after masking. The white stripes in the dynamic spectrum and butterfly diagram indicate the masked time bins during fitting of each considered burst. }
        \label{fig: frb ex2}
\end{figure*}

The header of the cutout dataset is partly inherited from the original dataset. There could be missing or wrong entries that need to be properly fixed, and this is done in the logfile to avoid manipulating the rigid filterbank-format structure. In this section, we focus on some caveats regarding the NC--FRB data.

The recorded observation start time reported in the header of the cutout dataset is the one of the original dataset. For independent calculation of the absolute ToA, which would be otherwise not possible, a keyword indicating the cutout starting time is added in the logfile. For 4 out of 29 datasets the starting time was not correctly written in the header due to technical issues during the data acquisition. In such cases, the observation date was fixed in the logfile, but the absolute ToA of the FRB cannot be properly recovered, therefore a warning note and a value of ${\rm ToA}=-1$ are reported.

The current version (v. 1.0) of {\tt archivist}  performs the fitting of the burst profile with a single Gaussian function. However, the shape of FRB profiles may depart from a simple Gaussian function, exhibiting for example decaying tails or secondary bursts (e.g. \citealt{Thornton13,Champion16,Zhu20,Jankowski23}). In general, the Gaussian profile is an adequate representation of the NC--FRB profiles observed to date, as also indicated by the typical $\chi_{\rm red}^2 \sim 1$. Nonetheless, as shown in Figure \ref{fig: frb ex3} as an example, the profile of FRB20220912A observed on 22-Aug.-2023 exhibits a decaying tail. In this case, the Gaussian profile is not able to fully reproduce the tail, yielding a worse $\chi_{\rm red}^2 \sim 1.74$. More complex fitting functions could become necessary as the number of bursts detected by the NC with non-Gaussian morphologies increase. Accordingly, future releases of {\tt archivist} may include alternative models to better reproduce the observed profiles.

When dealing with sub-bursts, {\tt archivist} includes the possibility of temporal masking, which reduces contamination and improves the fit goodness of a specific burst. As a representative example, we reported in Figure \ref{fig: frb ex2} the case of FRB20240114A (observation of 17-Mar.-2024), which shows multiple close sub-bursts. We fitted the two brightest bursts separately with a Gaussian function, while masking the other peaks to improve the fidelity of the fit. The two secondary bursts remain currently unmodelled in our analysis. Although not implemented in {\tt archivist}, a possible strategy for dealing with close sub-bursts is the simultaneous fitting of multiple Gaussian components.

We stress that the FRB shape depends on the temporal resolution. While the performed temporal downsampling improves the burst SNR, possible sub-structures might be smoothed.  Studying the burst profiles at high resolution is beyond the scope of the present work, but this will be possible for {\tt INCART} users thanks to the availability of cutout data saved at the temporal resolution of $\tau_{\rm samp}=138 \; {\rm \mu s}$.

\section{Discussion}
\label{sect: Conclusions}

\begin{figure}
        \centering
\includegraphics[width=0.45\textwidth]{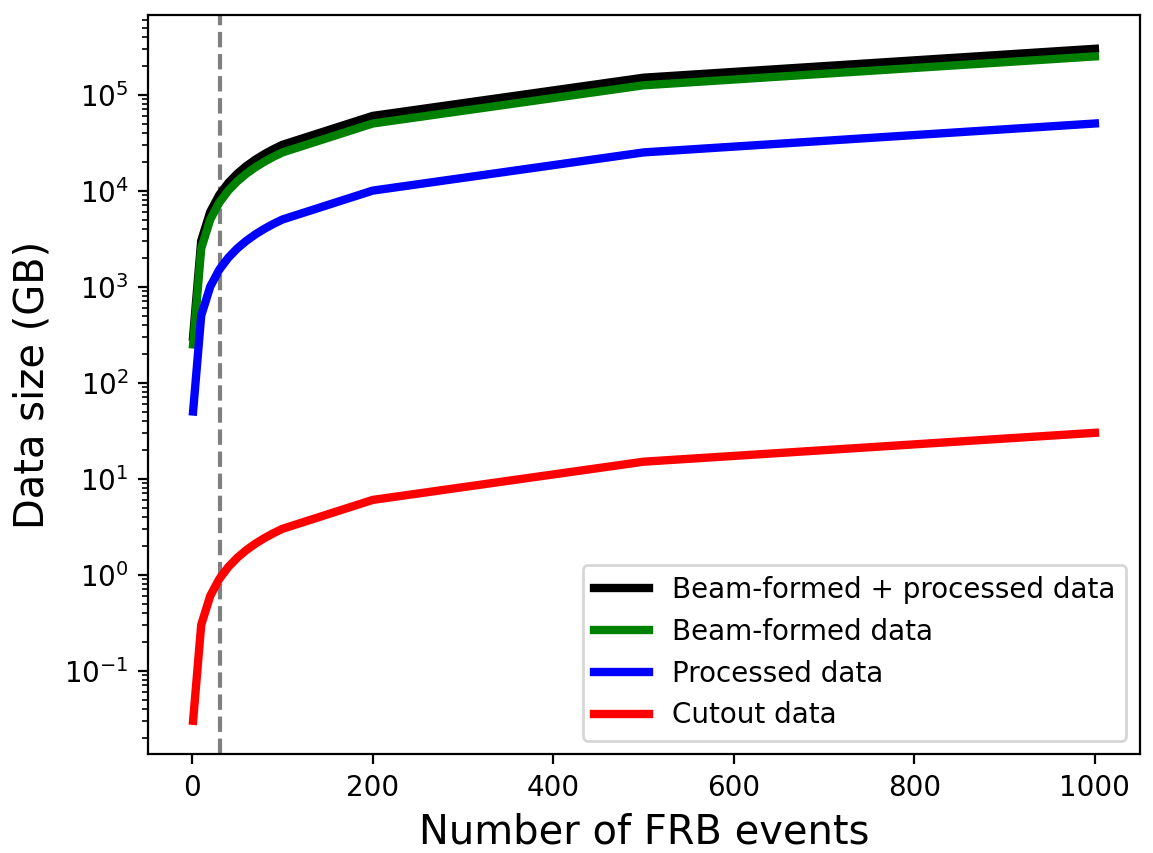}
        \caption{Expected data size with increasing number of FRB detections from the NC. Curves report data sizes for total beam-formed plus processed (black), beam-formed (green), processed (blue), and cutout (red) data based on typical single data size of present NC observations. The vertical line refers to the  number of FRBs detected as of the end of 2025.}
        \label{fig: data size}
\end{figure}

Data storage is nowadays a key challenge of all radio observatories. Depending on the data format, several techniques have been developed to efficiently reduce the data size. The data stored in {\tt INCART} are in filterbank format and consist of short cut outs (a few seconds) around the burst. This procedure of temporal extraction from the entire observation allows us to save a remarkably large amount of disk space, as shown in Figure \ref{fig: data size}. Based on the current data rate (see Section \ref{sect: Data processing}), we estimate that with 900 FRB events (a factor of 30 higher than the present NC detections), the beam-formed plus processed filterbank datasets would occupy a volume of $\sim 260$ TB (with the processed data being $\sim 40$ TB). Such number of detections are expected to be reached within $\sim 6$ yr after the NC upgrade completion (\citealt{locatelli20}). The cutout strategy, with consequent removal of beam-formed and processed datasets, reduces the data size by a factor of $\sim 10^4$, thus optimising the data volume for long-term storage.

Different solutions have been adopted by other international FRB detectors for data access, which is typically limited to short fractions of the total observing time to contain the large data volume. For a comparison with {\tt INCART}, we provide an overview of the services offered by some facilities. The Canadian Hydrogen Intensity Mapping Experiment (CHIME; \citealt{Bandura14,CHIMECollaboration22_TELESCOPE}) provides open access both to a database\footnote{\url{https://www.chime-frb.ca}} presenting derived physical quantities and to the Canadian Advanced Network for Astronomical Research (CANFAR\footnote{\url{https://www.canfar.net/storage/list/AstroDataCitationDOI/CISTI.CANFAR/}}) collecting dynamic spectra in hierarchical data format (HDF5) for diagnostic inspection. For some events, HDF5 advanced data products are also made public for independent re-analysis \citep{Amiri24}. The MeerTRAP project \citep{Sanidas18} performs a real-time, commensal transient search with MeerKAT and reports discovery products via publications and the SARAO\footnote{\url{https://archive.sarao.ac.za}} archive. However, a public database similar to those by CHIME and the NC has not yet been announced. The Australian Square Kilometre Array Pathfinder (ASKAP) monitoring through the Commensal Real-time ASKAP Fast Transients (CRAFT; \citealt{Scott25}) survey provides a public archive\footnote{\url{https://researchdata.edu.au/craft-high-time-release-1}} with dynamic spectra as NumPy-format arrays and diagnostic products. For selected events, raw voltages and processed data enabling independent re-analysis are also available. Finally, \cite{Zhang20_PARKESdb} developed a database tailored for the Parkes radio telescope.  Specifically, the stored data products include small data segments containing single pulse signals (both candidate transient phenomena and RFI) obtained from reprocessing of archival observations with advanced homogeneous methods. In summary, regardless of the variety of formats for the data products, {\tt INCART} aligns with the strategies adopted by other services, aiming at limiting the data volume and offering manageable datasets.

\section{Summary and conclusions}
\label{sect: Conclusions2}

In this work, we presented {\tt INCART}, a public online repository that provides the community with access to advanced data products of FRB events detected by the NC radio telescope. Specifically, {\tt INCART} collects the catalogue of NC-FRBs, which can be consulted online or downloaded in FITS format, and manageable processed datasets in filterbank format suitable for independent analysis of the FRB properties. As the physical parameters were derived through homogenous procedures (Sections \ref{sect: Re-analysis steps}, \ref{sect: Equazioni}) that might not be optimal for complex bursts, such independent analysis may include, for instance, fitting the FRB with alternative profiles beyond a simple Gaussian, or investigating possible sub-structure at high temporal resolution (Section \ref{sect: Caveats}). Moreover, the usage of the NC-FRB  data products in synergy with those from other detectors enables, for example, multi-epoch and multi-frequency studies of FRBs.

The large data volume to be stored, which is a common challenge for all radio facilities and especially FRB detectors observing at high temporal resolution, is handled by saving in {\tt INCART} only short time segments of interest. This strategy guarantees the optimisation of the disk space for long-term storage (Section \ref{sect: Conclusions}). Furthermore, in the view of the planned transient real-time buffer \citep{Naldi25}, improved real-time searching algorithms are under development \citep{Camilleri26}, which will also reduce the short-term data size.

In conclusion, the scalable design of {\tt INCART} will also ensure access to the FRB data products of the NG-Croce in the next years. The incoming upgrade of the NC will significantly improve the instrumental capabilities, allowing the exploitation of all N-S cylinders, the E-W arm, and the multi-beam observing mode. These will lead to improvements in sensitivity, field of view, and localisation, further enhancing the scientific role of the NC-FRB project and the value of its data products for the community.

\bibliographystyle{aa}
\bibliography{bibliografiaFRB}

\begin{acknowledgements}
The authors thank the referee for carefully reading our manuscript and providing valuable comments and suggestions. LB acknowledges F. Bedosti, G. Lorenzo, A. Magro, and N. Ragno for IT support and suggestions. The research activities described in this paper were carried out with contribution of the NextGenerationEU funds within the National Recovery and Resilience Plan (PNRR), Mission 4 - Education and Research, Component 2 - From Research to Business (M4C2), Investment Line 3.1 - Strengthening and creation of Research Infrastructures, Project IR0000026 – Next Generation Croce del Nord. This research made use Astropy, a community-developed core Python package for Astronomy \citep{astropycollaboration13,astropycollaboration18}, Matplotlib \citep{hunter07MATPLOTLIB}, NumPy \citep{harris20NUMPY}, SciPy \citep{scipy}. This research made use of SigPyProc (\url{https://github.com/FRBs/sigpyproc3}), a Python package for FRB and pulsar data analysis.

\end{acknowledgements}

\section*{Data Availability}

The data and corresponding products (catalogues, diagnostic plots, logfiles) described in this work can be retrieved via {\tt INCART}. The same data are also accessible via the INAF Radio Data Archive. A simplified version of the {\tt archivist} pipeline performing key steps of the FRB analysis chain is under development, and will soon become available to be shared upon request.

\end{document}